\documentclass[11pt]{article}

\usepackage{natbib}
\usepackage{graphicx}
\usepackage{hyperref}
\usepackage{amsmath}
\usepackage{amssymb}
\usepackage{placeins}
\usepackage{lmodern} 
\usepackage{lineno}
\newcommand{\rsc}{r_{\mathrm{sc}}}

\newcommand{\rzero}{r_{0}}
\newcommand{\gzero}{g_{0}}
\newcommand{\mDM}{m_{\mathrm{DM}}}
\newcommand{\Mbar}{M_{\mathrm{bar}}}
\newcommand{\DMrat}{\mDM/\Mbar}
\newcommand{\mrat}{m_{\mathrm{rat}}}
\newcommand{\gbar}{g_{\mathrm{bar}}}
\newcommand{\gobs}{g_{\mathrm{obs}}}

\newcommand{\gDM}{g_{\mathrm{DM}}}
\newcommand{\vobs}{v_{\mathrm{obs}}}

\newcommand{\vu}{v_{\mathrm{u}}}

\newcommand{\rhoDM}{\rho_{\mathrm{DM}}}

\newcommand{\gunit}{10^{-12} \ {\mathrm{m/s}}^2}

\title{A Common Galaxy-Independent Radial Structure of Dark Matter}

\author{%
P. Steffen\textsuperscript{1$\star$} \\
\\
\textsuperscript{1}\normalsize Deutsches Elektronen-Synchrotron DESY, Hamburg, Germany \\
\textsuperscript{$\star$}\normalsize corresponding author: peter.steffen@desy.de \\
}

\date{\ }

\begin{document}

\maketitle

\begin{abstract}

A galaxy-independent scaled radial coordinate is introduced, 
extending from the galactic center to the radius where the baryonic acceleration equals $10^{-12}\ m/s^2$.
Using this coordinate, the SPARC galaxies are found to share a common radial structure,
resulting in common radial distributions of the enclosed dark-matter mass, the dark-matter density, and the unified velocity.
These quantities constitute mathematically equivalent representations of the common radial structure.
Consequently, the radial acceleration relation, the approximately flat outer rotation curves, and the BTFR 
follow directly from the fitted radial dependence of the enclosed dark-matter mass.

The enclosed dark-matter mass itself increases  linearly with the scaled radius,
 reaching about seven times the enclosed baryonic mass within the observed SPARC range.

The common radial structure reported here is determined from the SPARC sample,
because it provides sufficiently accurate rotation curves to determine the radial dark-matter distribution empirically.
However, there is presently no observational evidence that the derived  common radial structure is unique to the SPARC sample.
The common radial structure identified here 
provides new empirical information about the organization of dark matter in galaxies. 
Beyond serving as an observational constraint on halo models, 
it may also help guide the development of a deeper physical understanding of the processes responsible for this organization.

\end{abstract}

\newpage 
\section{Introduction}
\label{sec:1}

The radial acceleration relation (RAR) demonstrated that the observed acceleration $\gobs$
 is closely correlated with the baryonic acceleration $\gbar$ 
over a wide range of galaxies \cite{McGaugh2016}.
Since the baryonic acceleration is related to the radial distance through Newtonian gravity,
\begin{equation}
\gbar = \frac{G\ \Mbar}{r^2}.  \label{equ:Newton}
\end{equation}
the RAR naturally raises the question whether it also contains information about a common radial structure of galactic dark matter.

Direct comparisons of radial distributions between galaxies are difficult 
because galaxies differ substantially in mass and size. 
However, Newtonian gravity immediately suggests a natural scaling,
\begin{equation}
r \propto  \frac{1}{\sqrt{\gbar}}
\end{equation}
which removes the trivial dependence on baryonic mass and provides a common radial coordinate for all galaxies.

This work investigates whether the SPARC rotation-curve data reveal a common radial structure when expressed in this scaled coordinate.
 The analysis is entirely empirical and does not assume a parametric halo model.
 Instead, it derives the enclosed dark-matter fraction directly from the observed acceleration relation and examines its radial behavior.

The resulting radial representation reveals a remarkably simple common radial structure.
 The enclosed dark-matter fraction increases approximately linearly with scaled radius, 
implying an $r^{-2}$  density profile and a unified velocity profile. 
These relations provide a compact radial description of the SPARC observations
 and establish direct quantitative connections between the enclosed dark-matter mass, 
the density distribution, the observed acceleration, the approximately flat outer rotation curves, and the BTFR.

Although extensive halo-model analyses of the SPARC sample (e.g. \cite{Li2018},  \cite{Li2020},  \cite{Li2022})
have greatly improved our understanding of galaxy halos, 
no single model has emerged as universally preferred. 
This work therefore complements these studies by focusing on the  common radial structure 
directly implied by the observations.

\section{Natural Radial Coordinates}
\label{radial_ansatz}

 Equation \ref{equ:Newton} yields
\begin{equation}
r = \frac{1}{\sqrt{\gbar}}\ \  \sqrt{G\ \Mbar}.  \label{equ:r}
\end{equation}
A basic acceleration $\gzero = \gunit$ as reference gives
\begin{equation}
r = \frac{1}{\sqrt{\gbar/\gzero}}\ \  \sqrt{\frac{G\ \Mbar}{\gzero}}.  \label{equ:Nr}
\end{equation}

The first factor, $1 / \sqrt{\gbar/\gzero} = \rsc$, describes a scaled radial dependence, 
while the second factor, $\sqrt{G\ \Mbar / \gzero} = \rzero$, describes the galactic scale,
 such that
\begin{equation}
r = \rsc \  \rzero     	 \label{equ:rsc_zero}
\end{equation}

The galactic scale,  $\rzero$, is the radial distance from the center, where $\gbar = \gunit$.
The scaled radius, $\rsc$,  is a dimensionless linear radial distance ranging  from zero, at the galactic center,
to 1.0 at $\gbar = \gunit$.

\section{Empirical Radial Dependence of the DM Mass Ratio}
\label{sec:rscdep}

\subsection{Determination of the DM mass ratio}
\label{rsc_fit}

The results of 2693 rotation-curve measurements from 153 galaxies are freely available from the \citep{SPARC}.
We use the measured baryonic acceleration $\gbar$ and the observed centripetal acceleration $\gobs$, 
together with their uncertainties, to determine the mass ratio as a function of the $\rsc$.

$\gobs$ and $\gbar$ are correlated over a wide range of galaxies \cite{McGaugh2016}.
This correlation can be expressed using $\rsc = 1 / \sqrt{\gbar/\gzero} $ as variable.

 The direct fit of  $\gobs = \gobs(\rsc)$ is not pursued here 
because the enclosed dark-matter mass ratio combines the measured quantities $\gobs$ and $\gbar$ 
into the quantity of physical interest before the fit is performed.
\begin{equation}
DM_{rat} = \frac{\mDM}{\Mbar}   =   \frac{\gDM}{\gbar}   =    \frac{\gobs}{\gbar}  - 1  \ ,	 \label{equ:DMrat}
\end{equation}
 based on the relation $\gDM = \gobs - \gbar$,
where $\gDM (\rsc)$ is the acceleration by the enclosed dark-matter mass, $\mDM$.
Here the uncertainties are combined in the fraction $\gobs/\gbar$.

The original data,  given in logarithms of the acceleration, 
 are converted to accelerations in units of $\gunit$.
The quoted uncertainties are given in dex for the logarithmic accelerations and are converted accordingly to relative uncertainties of the linear accelerations.
Because this analysis is based on linear combinations and differences of accelerations 
(e.g. $\gDM = \gobs - \gbar$), 
the logarithmic SPARC accelerations are converted to linear units before performing the analysis.

A least-squares fit  to the data gives the radial behavior of 
\begin{equation}
DM_{rat} (\rsc) = (10.42 \pm 0.15)\  \rsc  - ( 0.35 \pm 0.05) .			\label{equ:mratfit}
\end{equation}
A weighting of the data has not been used because the relative errors of 
$\gobs$ and $\gbar$ do not fluctuate much.
The uncertainties of the fitted parameters reflect the observed fluctuations of the DMrat values 
rather than the quoted relative measurement uncertainties

The $DM_{rat}$ values are shown  in the top panel of Figure \ref{fig:SPARC1} as a function of $ \rsc$ as black dots.
The fit is shown as a red diagonal line.
The data are well described by  the linear increase  of the mass ratio as a function of $\rsc$,
reaching  $DM_{rat}  \approx 7$ at $\rsc = 0.7$.

The data, used by the fit, are selected to be in the range   $ 0.064 < \rsc < 0.785 $ 
indicated by the two vertical red  lines in the figure.
Below the lower limit of   $\rsc  =  0.064$, there are no significant values of $DM_{rat}$,
above  the upper limit of  $\rsc  =  0.785$, there are nearly no data available.
 
The fitted line can be extrapolated to $\rsc =  0.033$, where $DM_{rat}  = 0$, 
a kind of threshold of $DM_{rat}$.  
However, one cannot exclude that the linear curve flattens out leaving some dark-matter mass for the 
inner galactic volume.

\subsection{Quality of the Fit}
\label{rsc_Qfit}

The quality of the fit result is evaluated by an analysis of the relative residuals:
\begin{equation}
\delta \mrat = \frac{\mrat(meas) - \mrat(fit)}{\mrat(fit) }.
\end{equation}
They have a mean of 
\begin{equation}
\overline{\delta \mrat} =  - 0.015  \pm 0.009 .  	\label{equ:mfitres}
\end{equation}
compatible with zero.
The fluctuation of the residuals is described by the standard deviation 
\begin{equation}
STD =  0.39 ,
\end{equation}
to be compared with the mean relative error  of $\overline{\sigma(\gobs/\gbar) /  (\gobs/\gbar)}= 0.31$.
A full decomposition of the residual scatter into distance, inclination, and mass-to-light-ratio systematics is beyond the scope of this empirical analysis.

\begin{figure*}[h]
\begin{center}
%\vspace{-3cm}
\includegraphics[width=\textwidth,height=0.80\textheight]{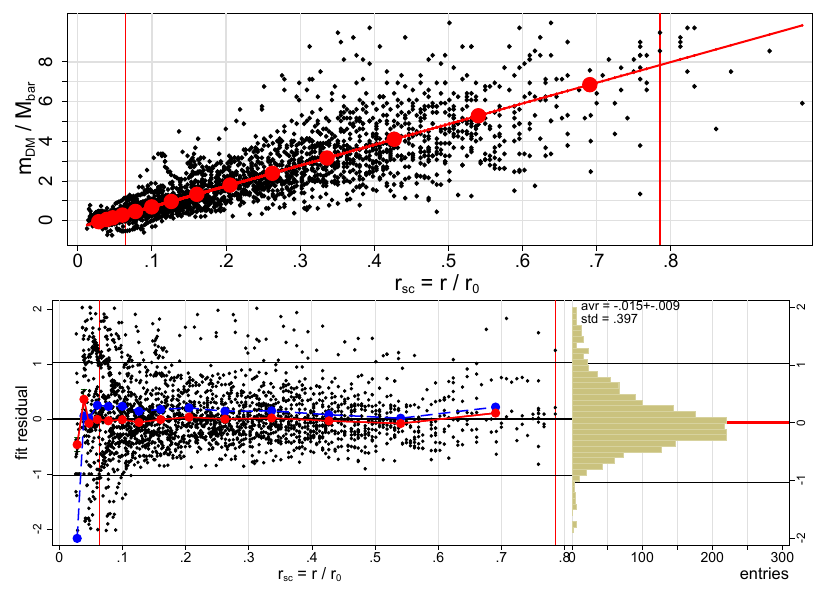}
\caption{The fit to the $DM_{rat}$ values as a function of $\rsc$.\\
top panel: 
Data and fitted curve. 
Averages of data within $\rsc$-bins are shown as thick red points. Further details are within the text.\\
bottom panel: 
Fit residuals and projection onto the y-axis.
Average residuals  of data within $\rsc$-bins are shown as thick red points, errors are within the point size.
Further details are within the text.
}
\label{fig:SPARC1}
\end{center}
\end{figure*}

The bottom panel of Figure \ref{fig:SPARC1} shows the residuals as a function of $\rsc$
as well as the projection onto the y-axis.

Average residuals  of data within $\rsc$-bins are shown as thick red points. 
The statistical errors are within the point size.
They demonstrate  the absence of systematic deviations as a function of $\rsc$.
Deviations are visible only for  $\rsc < 0.05$,
where the uncertainty of the mass ratio is large.
Measurements with residuals exceeding 2.5 standard deviation are eliminated for the final fit,
shown as black horizontal lines.
The  data range is indicated by the two vertical red lines.

For comparison thick blue points show the average residuals of  the fit by \citep{McGaugh2016} 
(converted to $DM_{rat} ( \rsc)$):
systematic deviations from zero are visible for the lower values of $\rsc$,
resulting in a mean residual of $0.131 \pm 0.01$
The origin of this systematic trend is not investigated further here. 
It may reflect the adopted one-parameter functional form of the empirical RAR relation

In summary, the  relation $\DMrat(\rsc)$ of Equation \ref{equ:mratfit} 
 provides a consistent and uniform description of the SPARC data
 within the quoted acceleration uncertainties.
No systematic trend is seen in the residuals over the investigated acceleration range; 
possible dependencies on galaxy subclasses are not tested separately here.
It  provides a robust  basis for the derivation of a  galaxy-independent radial structure.
This formulation does not introduce new empirical information beyond the acceleration relation itself,
 but provides a direct radial representation that makes the implied halo structure explicit.

%\clearpage 

\clearpage

\section{Consequences of the Fitted Radial Dependence}

\begin{figure*}[h]
\begin{center}
%\vspace{-3cm}
\includegraphics[width=\textwidth,height=0.75\textheight]{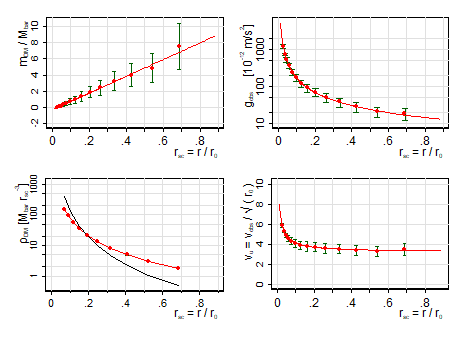}
\caption{
Radial dependencies derived from the fitted DM mass ratio.
 The upper-left panel shows the fitted dark-matter mass ratio $\DMrat$.
The remaining panels show quantities derived from this relation: 
the observed acceleration $\gobs$ (upper right), 
the dark-matter density profile $\rhoDM$ (lower left), 
and the unified velocity $\vu$  (lower right). 
The red curves show the functional form of the quantities.
The red points with error bars show the averages and standard deviations of different  $\rsc$-bins.}
\label{fig:summary}
\end{center}
\end{figure*}

Figure \ref{fig:summary} shows the fit result, Equation \ref{equ:mratfit},  in the upper right panel 
together with the radial function of related quantities.
These are:
\begin{enumerate}
\item
{\bf RAR relation  $\gobs(\rsc)$}
 The  common, galaxy independent radial distribution is obtained from Equation \ref{equ:DMrat} : 
\begin{equation}
\gobs(\rsc)     	 = ( gDM (\rsc) + 1) \  \gbar = ( gDM (\rsc) + 1)  / \rsc^2 .
\end{equation}
Substituting Equation \ref{equ:mratfit} yields 
%{10.42 \pm 0.15} \rsc  - ( 0.35 \pm 0.05)
\begin{equation}
\gobs(\rsc) = \frac{10.42 \pm 0.15}{\rsc}  + \frac{ 0.66 \pm 0.05}{\rsc^2} .			\label{equ:grscfit}
\end{equation}
Figure \ref{fig:summary} shows the function  in the upper right panel.
\item
{\bf DM Mass Density}
It is  the differential quotient 
\begin{equation}
\rhoDM (\rsc)   =  \frac{d\  \DMrat (\rsc) }{d\ V(\rsc)}    =     \frac{d\  \DMrat (\rsc) / d\ \rsc }{d\ V(\rsc) / d\ \rsc}               
\end{equation}
The numerator, $d\  \DMrat (\rsc) / d\ \rsc =  (10.42 \pm 0.15) $, is obtained from Equation \ref{equ:mratfit},
while the denominator is the differential shell volume, $d\  V (\rsc)  / d\ \rsc = {4/3\ \pi \   d\  \rsc^3}  =  
  {4\ \pi \ \rsc^2 \   d\   \rsc} $.
The result is 
\begin{equation}
\rhoDM (\rsc)   =   (0.829 \pm  0.012) / \rsc^2 ,     \label{equ:rhoDM}   
\end{equation}
a $\rsc^{-2}$ dependency of the DM mass density.
Figure \ref{fig:summary} shows the function  in the lower left panel.

The derived $\rsc^{-2}$ dependence corresponds to an isothermal-like density profile 
\cite{BinneyTremaine2008} and is therefore consistent with approximately flat outer rotation curves. 
It differs from the asymptotic behavior of the commonly used NFW halo \cite{Navarro1997NFW}, which approaches 
$\rsc^{-1}$  toward the centre and $\rsc^{-3}$ at large radii, indicating 
that Equation \ref{equ:rhoDM} characterizes the radial range covered by the SPARC data.

\item
{\bf Radial Velocity}
It is is obtained from the observed acceleration:
\begin{equation}
\gobs = \frac{\vobs^2}{r} .
\end{equation}
 $ r = \rzero \cdot \rsc$  yields
\begin{equation}
\vobs =\sqrt{\gobs\ \rsc}\  \sqrt{\rzero} = \vu \ \sqrt{\rzero},      \label{equ:vobs}
\end{equation}
where $\vu(\rsc)$ gives the radial shape of the radial distribution while $\rzero$ gives the galactic scale of it.
 Equation \ref{equ:grscfit} yields 
\begin{equation}
\vu(\rsc) = \sqrt{\gobs\ \rsc} = \sqrt{(10.42 \pm 0.15)   + \frac{ 0.66 \pm 0.05}{\rsc}  } .  \label{equ:vu}
\end{equation}
Figure \ref{fig:summary} shows the function  in the lower right panel.

\end{enumerate}
In general for a selected galaxy, these common, galaxy-independent radial distributions have to be scaled by $\rzero$ 
in order to obtain the galaxy-specific r-distributions.
 
\clearpage

\clearpage

\section{Discussion}

\subsection{ Connection to the  RAR, Flat Rotation Curves, and the BTFR}

The fitted dark-matter mass ratio of Equation \ref{equ:mratfit} provides the  basis 
from which several established  relations can be derived.
These are:
\begin{enumerate}
\item
{\bf RAR} (\cite{McGaugh2016})
It is the relation between the observed and expected acceleration of the SPARC data.
Equations \ref{equ:grscfit} is converted to the usual $\gbar$-dependence:
\begin{equation}
\gobs = (10.42 \pm 0.15)\ \sqrt{\gbar} +  ( 0.66 \pm 0.05)  \gbar   		\label{equ:RAR}
\end{equation}
which reproduces the observed radial acceleration relation.

\item
{\bf Flatness of the rotation velocity}
The unified velocity is described by Equation \ref{equ:vu} as a function of the scaled radial distance $\rsc$.
It is shown in  the lower right panel of Figure \ref{fig:summary}.
It shows a steep fall for $\rsc < 0.15$. 
For $\rsc > 0.2 $ rsp $\gbar < 24\  \gunit$ the unified velocity varies only from about 3.7 to 3.4, 
corresponding to $\vu = 3.55 \pm 0.15$.
This naturally explains the approximately flat outer rotation curves first reported by \citep{Rubin1980}.

The observed velocity, $\vobs = \vu \ \sqrt{\rzero}$,  is the unified velocity scaled by  
 $\rzero = \sqrt{G\ \Mbar / \gzero} $.
It  is a galaxy-specific scale that fixes the level of the $\vobs$ distribution, without changing the shape 
including the flatness.

\item 
{\bf BTFR} (\cite{Tully1977}, \cite{McGaugh2000},  \cite{Verheijen2001})
is the relation between  the observed velocity,  $\vobs$,  and the galactic baryonic mass $\Mbar$.
Using  $\rzero =\sqrt{G\ \Mbar / \gzero}$  in 
Equation \ref{equ:vobs}, $\vobs = \vu \ \sqrt{\rzero}$,  gives
\begin{equation}
\vobs^2 = \vu^2 \ \sqrt{G\ \Mbar / \gzero}      
\end{equation} 
and
\begin{equation}
\vobs^4 = \vu^4  \frac{G}{ \gzero}    \Mbar  
\end{equation} 
The approximately constant value of $\vu$  in the low-acceleration regime 
implies the approximate relation
\begin{equation}
 \Mbar  \propto   \vobs^4  
\end{equation} 
in the range of $\gbar \lesssim  24\  \gunit $.
The variation of $\vu$ over the observed low-acceleration range 
corresponds to an average variation of $\Mbar \ \mbox{by} \ \pm 17\%$. 
\end{enumerate}

\subsection{Outlook}

The  SPARC data establish the approximately linear growth of the enclosed dark-matter fraction 
over the range $ 0.064 < \rsc < 0.785 $ . 
However, the statistical significance of the data 
decreases considerably in the range $\rsc > 0.5$. 
Therefore the linear growth of $\DMrat$ may deviate from linearity and approach saturation.

 Additional high-quality rotation-curve measurements at both smaller and larger scaled radii would be valuable. 
Data at $\rsc < 0.06$ could clarify the onset of the dark-matter contribution and
 test for a threshold or flattening of the relation.

Data at $\rsc > 0.8$  could determine whether the observed linear growth continues 
or approaches saturation at larger radii.

\subsection{Implications of the Common Radial Structure}

Our analysis shows that the enclosed dark-matter fraction
 increases  linearly throughout the radial range covered by the SPARC data, 
reaching $\DMrat \approx 7$ near the largest observed scaled radii. 
This implies that a substantial fraction of the enclosed mass is located in the outer parts of the galaxies sampled by SPARC.

Although the behavior beyond the observed range is presently unknown, 
any successful model of dark matter or galaxy formation should reproduce this empirical radial growth over the measured range. 
Future rotation-curve measurements extending to larger scaled radii could determine whether the growth continues or approaches saturation.

The linear increase of the enclosed dark-matter mass with $\rsc$
 implies $\rhoDM(\rsc) \propto \rsc^{-2}$.
 Conversely, an approximately $\rsc^{-2}$ density profile produces linear enclosed-mass growth over the same radial range. 
The empirical slope and offset of Equation \ref{equ:mratfit} are not fixed by the $\rsc^{-2}$  radial dependence and therefore provide additional observational constraints on halo models. 
Equation \ref{equ:DMrat} shows  that the linear mass relation also reproduces the empirical acceleration relation. 
Together, the mass ratio, density profile, and acceleration relation represent different descriptions of the same common radial structure.

\section{Summary}

Expressed in a galaxy-independent scaled radial coordinate, the SPARC galaxies reveal a common radial structure.
 The enclosed dark-matter mass increases approximately linearly with the scaled radius, 
reaching about seven times the enclosed baryonic mass within the observed SPARC range. 
This empirical relation implies an $r^{-2}$ 
 dark-matter density profile and a unified velocity profile, and directly reproduces
 the radial acceleration relation, the approximately flat outer rotation curves, and the BTFR.

The scaled radial coordinate introduced here is defined independently of the SPARC sample.
Whether this common radial structure represents a general property of disk galaxies 
or extends to other galaxy populations remains an observational question. 
If confirmed by future observations, it may provide a useful empirical framework 
for understanding the organization of dark matter in galaxies.

\section*{Data availability}

The data underlying this article are publicly available from the SPARC database at
\url{https://astroweb.case.edu/SPARC/}.

\

\section*{Acknowledgment}

I thank the DESY Directorate and the IT division for their continuous support.  
I am grateful for the productive discussions with the GRAVI group at DESY,  
G.~Schierholz, U.~Martyn, and K.~Schmidt-Hoberg.  
Special thanks go to the SPARC collaboration for providing the astronomical data on which this analysis is based.

\bibliographystyle{mnras}
\bibliography{references}

\end{document}